\documentclass[aps,prd,secnumarabic,amssymb, amsmath,nobibnotes,nofootinbib,11pt]{revtex4}
\usepackage{amsfonts,amsmath,hyperref,url, color}
\usepackage{bm, bbm}
\usepackage{graphicx}
\usepackage{mathtools}

\newcommand{\be}{\begin{equation}}
\newcommand{\bal}{\begin{align}}
\newcommand{\eal}{\end{align}}
\newcommand{\ee}{\end{equation}}
\newcommand{\bea}{\begin{eqnarray}}
\newcommand{\eea}{\end{eqnarray}}
\newcommand{\bit}{\begin{itemize}}
\newcommand{\eit}{\end{itemize}}
\newcommand{\la}{\langle}
\newcommand{\ra}{\rangle}

\usepackage{color}

\begin{document}

\title{Horizon temperature on the real line}
\author{Michele Arzano}
\email{michele.arzano@roma1.infn.it}
\affiliation{Dipartimento di Fisica ``E. Pancini", Universit\`a di Napoli Federico II, I-80125 Napoli, Italy\\}
\affiliation{Dipartimento di Fisica,  ``Sapienza" Universit\`a di Roma, P.le A. Moro 2, 00185 Roma, Italy}
\author{Jerzy Kowalski-Glikman}
\email{jerzy.kowalski-glikman@ift.uni.wroc.pl}
\affiliation{Institute for Theoretical Physics, University of Wroc\l{}aw, Pl. Maxa Borna 9, Pl--50-204 Wroc\l{}aw, Poland\\}
\affiliation{National Centre for Nuclear Research, Ho˙za 69, 00-681 Warsaw, Poland}

\begin{abstract}
We illustrate the analogue of the Unruh effect for a quantum system on the real line. Our derivation relies solely on basic elements of representation theory of the group of affine transformations without a notion of time or metric. Our result shows that a thermal distribution naturally emerges in connecting quantum states belonging to representations associated to distinct notions of translational symmetry. 
\end{abstract}

\maketitle

The notion of horizon temperature is central for our understanding of the interplay between gravity and the quantum realm. Starting from the discovery that black holes radiate with a temperature proportional to their horizon surface gravity \cite{Hawking:1974sw}, over the years it was appreciated that such phenomenon is a universal feature for any local horizon \cite{Unruh:1976db,Gibbons:1977mu,Narnhofer:1996zk,Deser:1997ri,Jacobson:2003wv}. In particular, causal horizons can exist also in flat space for observers undergoing constant acceleration, and the associated temperature, proportional to their acceleration, was originally derived by Unruh \cite{Unruh:1976db}.

The thermodynamic properties of horizons seem to hold the key to a more fundamental understanding of gravity itself. A celebrated result by Jacobson \cite{Jacobson:1995ab} shows in fact that Einstein's equations can be derived assuming an entropy balance relation $\delta S = \delta E/ T_U$, between the variation of the local horizon entropy $\delta S$, the Unruh temperature $T_U$, and the energy flow across the horizon $\delta E$, and setting it equal to the variation of the horizon area described in terms of the focusing of null geodesics. At the same time the Hawking emission associated to the horizon entropy of a black hole has led to one of the most puzzling dilemmas that our quest for a theory of quantum gravity is expected to dissolve: the apparent impossibility to reconcile unitarity of quantum evolution with the description of the full evaporation of a black hole \cite{Hawking:1976ra,Hawking:1982dj,Susskind:1993if,tHooft:1996rdg,Mathur:2009hf,Almheiri:2012rt}.  

In this letter we show how the same type of relationship between horizon and temperature found for a quantum field in various space-time geometries admitting an event horizon, emerges already in a simple quantum system with translational symmetry living on the real line. The geometric input for our quantum model is reduced to the bare minimum and essentially boils down to a partition of the real line determined by the choice of translations generated by a dilation, acting only on the positive or negative half of the real line. The inclusion of dilations enlarges the abelian group of usual translations on the whole real line to the group of {\it affine transformations} \cite{VileK,Moses:1974ez}. The derivation we present is essentially group theoretic in nature: a thermal spectrum is naturally emerging when comparing two representations of the affine group which differ in the range of their ``space spectrum" due to the partition induced by the dilation transformation.

To understand the relevance for the affine group in the present context let us recall that elementary particles are irreducible representations of semi-direct product groups \cite{Wigner:1939cj, Woit}. These are comprised of an abelian subgroup of translations acted upon by a group of ``generalized" rotations. Irreducible representations of such groups are constructed from (one-dimensional) representations of the abelian subgroup \cite{Barut:1986dd} which are nothing but plane waves i.e. wavefunctions representing free particles.

In the case of a line partitioned into two, positive and negative, semi-lines one deals with two candidate translation generators, of which one, denoted by $P$ below is just the standard momentum, generating the real line translations, while another, denoted by $R$ generates translations on a semi-line. One can consider plane waves, and therefore one-particle states, which are eigenfunctions of both translations $P$ and $R$. The co-existence of these two candidate translation generators is the key to the ``linear" Unruh effect we derive.

The group of transformations generated by $P$ and $R$, also known in the mathematical literature as the ``$ax+b$'' group, is the group of affine transformation of the real line i.e., as its name suggests, the transformations of the form
\be\label{ax+b}
x\rightarrow ax+b\,,\,\,\,\,\, a\in \mathbb{R}^+\,,\,\, b\in \mathbb{R}\,.
\ee
Setting $a=1$ we obtain the subgroup of translations of the real line; the action of this subgroup is transitive since for any given $x$ all elements of $\mathbb{R}$ can be obtained from it by an appropriate $b$. The subgroup of dilations defined by $b=0$ is transitive only on $\mathbb{R}^+$ (or $\mathbb{R}^-$). This subgroup can be regarded as a group of translations of the real positive (negative) semi-line. Therefore the $ax+b$ group can be seen as a group comprised of two types of translations, one on the real line and the other on half of it (positive or negative). Denoting the group element as $g=(a,b)$ the group multiplication law for two elements $g_1=(a_1,b_1)$ and $g_2=(a_2,b_2)$ is given by
\be
g_1 g_2 = (a_1 a_2, a_1 b_2+b_1)\,.
\ee
The unitary irreducible representations of the $ax+b$ group are well known 
(see e.g. \cite{VileK, Moses:1974ez} and \cite{Gayral:2007hk,Milburn:2007jw,Bergeron:2013ika} for recent applications). There are just two such representations. One is realized on the Hilbert space of functions on the positive real line whose elements we denote with $|k\ra_+$. The actions of translations $T(b) = (1, b)$ and dilations $D(a) = (a, 0)$ on such Hilbert space are given by \cite{VileK}
\be
T(b) |k\ra_+ = e^{- i b k}\,  |k\ra_+\,,\qquad D(a) |k\ra_+ =  |a\, k\ra_+\,.
\ee
and these can be rewritten in terms of the generators $P$ and $R$ as 
\bea\label{tidk}
T(\alpha) |k\ra_+  &= e^{-i \alpha P}\,  |k\ra_+ = e^{-i \alpha k}\,  |k\ra_+ \\
D(\lambda) |k\ra_+ & = e^{-i \lambda R}\,  |k\ra_+ =  |e^{-\lambda}\, k\ra_+ \label{dk} \,.
\eea
Differentiating with respect to the transformation parameters and setting them to zero we obtain the action of the generators
\be\label{pk+}
P\, |k\ra_+ = k\, |k\ra_+
\ee
and
\be\label{rk+}
R\,  |k\ra_+ = - ik\frac{d}{dk}\, |k\ra_+\,,
\ee
with $k\in \mathbb{R}^+$. These actions lead to the commutator
\begin{equation}\label{axbalg}
  [P,R]=iP
\end{equation}
among the generators which defines the Lie algebra of the $ax+b$ group, the {\it simplest} non-abelian Lie algebra. The transformations \eqref{tidk} and \eqref{dk} can be seen as the one-dimensional counterparts of infinitesimal Poincar\'e transformations with translations generated by $T$ and the dilations $D$ playing the role of boosts generators. Following this analogy we will interpret the states $|k\ra_+$ belonging to the irreducible representation of the group with positive eigenvalues for the generator $P$, as ``$P$-particle states". Let us note in passing that, if $R$ is to be interpreted as a generator of translations, it should, like $P$, carry a dimension of inverse length, contrary to what eq.\ (\ref{axbalg}) indicates. To simplify the following expressions, we will keep this generator dimensionless, restoring its physical dimension, when we turn to the discussion of our results.

The action of translations and dilations on functions $\psi(k) = {}_+\la k| \psi\ra$ 
belonging to the $P$-particle Hilbert space is explicitely given by
\be\label{groupfunp}
T(\alpha)\,  \psi(k) = e^{i \alpha k}\,  \psi(k)\,,\,\,\,\,\,  D(\lambda)\,  \psi(k) = \psi(e^{\lambda} k)\,,
\ee
and the invariant inner product on such space is defined using the invariant momentum space measure $\frac{dk}{k}$
\be\label{innprod+}
\langle \psi | \psi' \rangle = \int^{\infty}_0 \frac{dk}{k}\, \bar{\psi}(k) \psi'(k) =  \int^{\infty}_0 \frac{dk}{k}\, \langle\psi| k\rangle_+ {}_+\langle k| \psi'\rangle\,.
\ee
The other unitary irreducible representation is realized on the Hilbert space of functions on the negative real line, whose basis kets we denote  $|k\ra_-$ on which the generators $P$ and $R$ act as in \eqref{pk+}, \eqref{rk+}, and with inner product given by
\be\label{innprod-}
\langle \psi | \psi' \rangle = \int^{0}_{-\infty} \frac{dk}{k}\, \bar{\psi}(k) \psi'(k) =  \int^{0}_{-\infty} \frac{dk}{k}\, \langle\psi| k\rangle_- {}_-\langle k| \psi'\rangle\,.
\ee

We can introduce functions on the coordinate space dual to the momentum space real line using a Fourier transform written in terms of a combination of the two irreducible representations $|k\ra_+$ and $|k\ra_-$ as \cite{Moses:1974ez}
\be\label{fieldx2}
\la x|\psi\ra = \frac{1}{\sqrt{2 \pi}} \int^{\infty}_{0}\, \frac{dk}{k}\, e^{ikx}\,  {}_+\langle k| \psi\rangle + \frac{1}{\sqrt{2 \pi}} \int_{-\infty}^{0}\, \frac{dk}{|k|}\, e^{ikx}\, {}_-\langle k| \psi\rangle
\ee
from which, using \eqref{groupfunp}, one obtains
\be\label{geoax}
T(\alpha)\, \psi(x) =  \psi(x + \alpha) \,,\quad  D(\lambda)\,  \psi(x) = \psi(e^{-\lambda} x)\,,
\ee
so that
\begin{equation}\label{pronx}
  P= -i\frac{d}{dx}\,,\quad R= ix\frac{d}{dx}\,,
\end{equation}
providing a coordinate representation of the algebra \eqref{axbalg}. Notice how from \eqref{geoax} it is evident that, while ordinary translations generated by $P$ span the whole real line, adopting $R$ as a generator of translations restricts the range of the coordinate space accessible by a finite translation to a half line.  From \eqref{fieldx2} we see that the coordinate representation wave-functions associated to $P$-particle states are given by ``positive momentum" or ``right-moving" plane waves
\be
\la x| k\ra_{\pm} =  \frac{1}{\sqrt{2 \pi}}\, e^{ikx}\,,\quad k\in \mathbb{R}^{\pm}\,,
\ee
which are the counterpart of the usual positive energy plane waves representing one-particle states in quantum field theory. In fact, what we have done so far has been just a toy version of what normally one does when writing an on-shell scalar field in terms of unitary irreducible representations of the Poincar\'e group. In our case the role of boosts transformations is played by dilations acting on the real line rather than on space-time. The analogy can be made more explicit if we change integration variable in the second term of \eqref{fieldx2} and write
\be\label{fieldx2a}
\la x|\psi\ra = \frac{1}{\sqrt{2 \pi}} \int^{\infty}_{0}\, \frac{dk}{k}\left( e^{ikx}\,  {}_+\langle k| \psi\rangle + e^{-ikx}\, {}_-\langle -k| \psi\rangle\right)\,.
\ee
For a real function $\la x|\psi\ra$ we must require that ${}_-\langle -k| \psi\rangle = ({}_+\langle k| \psi\rangle)^*$, and thus denoting ${}_+\langle k| \psi\rangle\equiv a^*(k)$ we can write
\be\label{psikx}
\psi(x)\equiv\la x|\psi\ra = \frac{1}{\sqrt{2 \pi}} \int^{\infty}_{0}\, \frac{dk}{k}\left( e^{-ikx}\, a(k)+e^{ikx}\, a^*(k) \right)\,,
\ee
an expression formally analogous to the well known expression for a real classical (on-shell) scalar field on Minkowski space.

We now introduce a set of states which diagonalize the dilation generator $R$. In analogy with the states $|k\ra_{\pm}$, which carry a charge (momentum) associated to the translational symmetry generated by $P$, these states will carry a charge associated with the {\it multiplicative} translational symmetry generated by $R$ on the positive half line $\mathbb{R}^+$. From representation theory \cite{VileK} we known that functions which diagonalize $R$ can be obtained via the Mellin transform of functions defined on the positive half line, in our case functions on the Hilbert space spanned by $|k\ra_+$ 
\be\label{omegamellpsi}
\la \omega | \psi\ra =  \frac{1}{\sqrt{2 \pi}} \int^{\infty}_0\, \frac{dk}{k}\, k^{-i \omega}\, {}_+\la k| \psi\ra\,,\qquad \omega\in\mathbb{R}\,,
\ee
from which we can write
\be\label{omk+}
\langle \omega  | k \ra_+ = \frac{1}{\sqrt{2 \pi}}\, k^{-i \omega}\,.
\ee
The relation \eqref{omegamellpsi} allows us to express an $|\omega\ra $ state as a superposition of $P$-particle sates 
\be\label{omegamell}
|\omega\ra = \int^{\infty}_0\, \frac{dk}{k}\, k^{i \omega}\, | k \rangle_+\,,
\ee
from which, using \eqref{rk+}, we can write the following actions of the $ax+b$ generators
\begin{equation}\label{pronomega}
  P\, |\omega\rangle=|\omega-i\rangle\,,\quad R\, |\omega\rangle=-\omega\, |\omega\rangle\,,
\end{equation}
showing that the $|\omega\rangle$ states diagonalize the operator $R$, as expected. Using the inverse Mellin transform and ignoring convergence issues we can use \eqref{omegamell} to write a $P$-particle state as a superposition of  $|\omega\rangle$ states
\be\label{invmetransk}
 |k\rangle_{+} = \int^{\infty}_{-\infty}\, d\omega\, \langle\omega |k \rangle_+\, |\omega\rangle = \frac{1}{\sqrt{2 \pi}} \int^{\infty}_{-\infty}\, d\omega\, k^{-i\omega}\, |\omega\rangle \,,\\\\\\\\\\\ \ \ \ \ \ \ k\in\mathbb{R}^{+}\,.
\ee
Notice that making use of the distributional identities
\be
 \int^{\infty}_{-\infty}\, d\omega\, k^{i\omega - 1}\, (k')^{-i \omega} = 2 \pi\, \delta(k-k')\,,
\ee
and
\be
 \int^{\infty}_0\, \frac{d k}{k}\, k^{-i(\omega - \omega')}\, = 2 \pi\, \delta(\omega - \omega')\,,
\ee
it can be easily checked that the {\it covariant} inner product for $|k\ra_+$ states
\be
{}_+\la k| k'\ra_+ = k\, \delta(k-k')
\ee
is compatible with standard inner product
\be\label{omegainner}
\la \omega|\omega'\ra = \delta(\omega-\omega')
\ee
for $|\omega\ra$ states. We now have all the tools to re-write the field $\psi(x)=\la x|\psi\ra$ in terms of $R$-eigenmodes $\la\omega|\psi\rangle$. We start by writing equation \eqref{fieldx2a} as
\be
\la x|\psi\ra = \frac{1}{\sqrt{2 \pi}} \int^{\infty}_{0}\, \frac{dk}{k}\, \left(e^{ikx}\,  {}_+\langle k| \psi\rangle + e^{-ikx}\, {}_+\langle k| \psi\rangle^*\right)\,,
\ee
and substitute in it the expressions for  ${}_{+}\la k|\psi\rangle$ given in \eqref{invmetransk}
\be\label{fieldx3}
\la x|\psi\ra = \frac{1}{2 \pi} \int^{\infty}_{0}\, \frac{dk}{k}\,\int^{\infty}_{-\infty}\, d\omega\,  \left( e^{ikx}\, k^{i\omega}\, \la \omega|\psi\rangle +  e^{- ikx}\,k^{- i\omega}\, \la \omega|\psi\rangle^* \right)\,.
\ee
With the help of the identity
\be
\int_0^\infty \frac{dk}{k}\, k^z \, e^{\pm ikx} = \Gamma(z)\, x^{-z} e^{\pm i\pi z/2}\,,
\ee
we can thus write 
\be\label{fieldx3a}
\la x|\psi\ra = \frac{1}{2 \pi} \int^{\infty}_{-\infty}\, d\omega\,  \left( x^{-i\omega}\, e^{-\frac{\pi \omega}{2}} \Gamma(i\omega) \la \omega|\psi\rangle +  x^{i\omega}\, e^{-\frac{\pi \omega}{2}} \Gamma(-i\omega)\, \la \omega|\psi\rangle^* \right)\,.
\ee
Such equation can be ``inverted" \cite{Moses:1974ez} to obtain an expansion of the function $ \la \omega|\psi\rangle$ in terms of coordinate eigenfunctions $x^{i\omega}$ of the dilation generator
\be\label{opsi}
 \la \omega|\psi\rangle =  - \frac{\omega}{2 \pi}\, \Gamma (-i \omega)\, e^{-\frac{\pi \omega}{2}} \int^{\infty}_{-\infty}\, \frac{dx}{x}\, x^{i\omega}  \la x|\psi\ra\,.
\ee
Let us now consider any {\it real} function restricted to the positive half line
\be
{}_+\la x| \psi\ra = \psi(x)\,,\,\,\,\,\,\,\, x\in\mathbb{R}^+\,.
\ee
We know that eigenfunctions of the $R$ operator $x^{-i\omega}$ form a complete set on the positive (or negative) half line and thus ${}_+\la x| \psi\ra$ can be expanded in terms of such functions, i.e. we must have ${}_+\la x| \omega\ra \sim x^{-i\omega}$. The normalization of such functions, however, is not fixed a priori. In our case, for example, we must make sure that it is consistent with the normalization of the inner product of $|\omega\ra$ states
\be
\la \omega|\omega'\ra = \delta(\omega-\omega')
\ee
we adopted at above. Since equation \eqref{opsi} above must hold also for functions restricted to the half line we see that the basis functions must be normalized as
\be\label{xomnorm}
{}_+\la x| \omega\ra = -\frac{e^{\frac{\pi \omega}{2}}}{\Gamma(-i\omega)\, \omega}\, x^{-i \omega}\,,\,\,\,\,\,\,\, x\in\mathbb{R}^+\,.
\ee
It can be easily checked that such rather unattractive normalization makes indeed sense by noticing that substituting \eqref{xomnorm} in the inverse of \eqref{fieldx2} for positive momenta
\be
{}_+\la k| \psi \ra = \frac{k}{\sqrt{2\pi}} \int^{\infty}_{-\infty}\,dx\, e^{-ikx}\, \la x|\psi\ra\,,\,\,\,\,\,\,\, k\in\mathbb{R}^+\,,
\ee
one re-obtains the expression \eqref{omk+} for ${}_+\la k| \omega\ra$.

We want now to consider eigenstates of the $R$-operator whose coordinate representation is a {\it right-moving plane wave} carrying a (positive) charge $\omega$ associated to the dilation symmetry (i.e. the analogue of Rindler one-particle state wave-function). In other words we look for states $|\omega \ra_R$ such that
\be
\la \xi | \omega\ra_R = \frac{1}{\sqrt{2\pi}}\, e^{i\omega \xi}
\ee
where the $\xi$-coordinate is such that the representation of the dilation operator is given by
\be
R= - i \frac{d}{d \xi}\,.
\ee
From the $x$-representation \eqref{pronx} of the symmetry generators it is easy to see that $x= e^{-\xi}$ from which
\be
\la \xi | \omega\ra_R = \frac{1}{\sqrt{2\pi}}\, e^{i\omega \xi} = \frac{1}{\sqrt{2\pi}}\, x^{-i\omega} = {}_+\la x| \omega\ra_R\,,
\ee
and thus, confronting \eqref{xomnorm} with \eqref{omk+} we derive 
the following transition amplitude between $P$ and $R$-particle states, i.e. the wave-function representing a $| \omega\ra_R$ particle in the $P$-representation
\be
{}_+\la k| \omega\ra_R = -\frac{\omega}{2 \pi}\,\Gamma(-i\omega)\, e^{-\frac{\pi \omega}{2}}\, k^{i\omega}\,.
\ee
The norm squared of such wave-function with respect to the $P$-particle Hilbert space inner product \eqref{innprod+} is given by
\be
 \int^{\infty}_0 \frac{dk}{k}\, {}_R\la\omega| k\rangle_+ {}_+\langle k| \omega'\rangle_R= \frac{\omega^2}{2\pi} |\Gamma(i\omega)|^2 e^{-\pi \omega} \delta(\omega-\omega')\,.
\ee
which, using the equality
\be
|\Gamma(i\omega)|^2 = \frac{\pi}{\omega \sinh(\pi \omega)}
\ee
leads to
\be\label{thermak}
 \int^{\infty}_0 \frac{dk}{k}\, {}_R\la\omega| k\rangle_+ {}_+\langle k| \omega'\rangle_R=\frac{\omega\, \delta(\omega-\omega')}{e^{2\pi \omega}-1} \,.
\ee
Since the normalization of P-particle states
\be
{}_+\la k| k'\ra_+ = k\, \delta(k-k')\,,
\ee
can be understood in terms of a probability density associated to the P-particle state $|k\ra_+$ of $k$ particles per unit volume
, equation \eqref{thermak} shows that an $| \omega\rangle_R$ state carries instead a density of $\frac{\omega}{e^{2\pi \omega}-1}$ $P$-particles per unit volume i.e. a thermal distribution at a temperature $\frac{1}{2\pi}$. This somewhat heuristic interpretation will be confirmed below where we show that ${}_+\la k| \omega\ra_R$ is nothing but a Bogolubov coefficient whose modulus corresponds to the expectation value of the $\omega$-number operator in the $P$-vacuum.

In order to recast the derivation of the thermal spectrum above in a form which parallels the usual field theoretic treatment we start by re-writing the expansion \eqref{fieldx3} as
\begin{align}\label{fieldx4a}
\la x|\psi\ra & =\frac{1}{2 \pi} \int^{\infty}_{0}\, {d\omega}\, x^{-i\omega}\Gamma(i\omega)\left( e^{-\pi\omega/2} \la \omega|\psi\rangle+ e^{\pi\omega/2} \la -\omega|\psi\rangle^* \right) \nonumber\\
&+\frac{1}{2 \pi} \int^{\infty}_{0}\, {d\omega}\, x^{i\omega}\Gamma(-i\omega)\left( e^{-\pi\omega/2} \la \omega|\psi\rangle^*+ e^{\pi\omega/2} \la -\omega|\psi\rangle \right)
\end{align}
We see that the right hand side of this expression has a form analogous to \eqref{psikx} with the $P$ eigenfunctions $e^{ikx}$ replaced by the $R$ eigenfunctions  $x^{i\omega}$ and thus we can write
\begin{equation}\label{psiomegax}
\psi(x) =\frac{1}{\sqrt{2 \pi}} \int^{\infty}_{0}\, \frac{d\omega}{\omega}\left( x^{-i\omega}\,  b(\omega) + x^{i\omega}\,  b^*(\omega)\right)
\end{equation}
with
\be\label{bomom}
b(\omega) = \frac{ \omega}{\sqrt{2 \pi}}\, \Gamma(i\omega)\left( e^{-\pi\omega/2} \la \omega|\psi\rangle+ e^{\pi\omega/2} \la -\omega|\psi\rangle^* \right)\,.
\ee
Now we can make use of the fact that $|\omega\ra$ states are just the Mellin transform of the $P$-particle states $| k \rangle_+$  \eqref{omegamell}

\be
|\omega\ra = \int^{\infty}_0\, \frac{dk}{k}\, k^{i \omega}\, | k \rangle_+
\ee
and thus
\be
\la \omega|\psi\rangle = \int^{\infty}_0\, \frac{dk}{k}\, k^{-i \omega}\,{}_+\la k|\psi\rangle = \int^{\infty}_0\, \frac{dk}{k}\, k^{-i \omega}\,a^*(k)
\ee
and
\be
\la -\omega|\psi\rangle^* = \int^{\infty}_0\, \frac{dk}{k}\, k^{-i \omega}\,{}_+\la k|\psi\rangle^* = \int^{\infty}_0\, \frac{dk}{k}\, k^{-i \omega}\,a(k)\,.
\ee
Substituting these expressions in \eqref{bomom} we obtain the Bogolubov transformation between the $a$ and $b$ expansion coefficients
\be\label{bogo1}
b(\omega) = \frac{ \omega}{\sqrt{2 \pi}}\, \Gamma(i\omega) \int^{\infty}_0\, \frac{dk}{k}\, k^{-i \omega}\,  \left( e^{-\pi\omega/2} a^*(k) + e^{\pi\omega/2} a(k) \right)\,.
\ee
Upon quantization the creation and annihilation operators associated to the $a$ coefficients satisfy the usual relation
\be\label{aadag}
 \left[\hat a(k),\hat a^\dag(k')\right] = k\,\delta(k-k')
\ee
which, using \eqref{bogo1}, leads to
\be\label{bbdag}
 \left[\hat b(\omega),\hat b^\dag(\omega')\right] = \omega\,\delta(\omega-\omega')\,,
\ee
for the creation and annihilation operators of $R$-particles. It can be easily checked that the expectation value of the number operator $\hat b^\dag(\omega) \hat b(\omega)$ in the vacuum state of the $a$-operators (the $P$-particle vacuum) gives exactly the thermal state on the right hand side of \eqref{thermak}. This straightforward derivation is carried out in many discussions of the Unruh effect in the literature and the interested reader can consult for example \cite{Arefeva:2013ahl} for an explicit calculation starting from the Bogolubov transformation \eqref{bogo1}.

In order to see how the arguments given above are related to the familiar Unruh effect let us consider the two-dimensional Minkowski and Rindler spaces. In the former, we have two commuting generators of translations, $T_0$, $T_1$, and the group generated by them acts transitively. Similarly, in the Rindler space case, we also have a transitive action of an abelian translation group, given by transformations which, from the Minkowski space perspective, are generated by boosts $N$ and dilation $D$ \cite{Arzano:2017opw}. All together these four generators form the two-dimensional Weyl-Poincar\'e algebra, with the nontrivial commutators being
\begin{align}
&[N,T_0] = iT_1\,\quad [N,T_1] = iT_0\nonumber\\
&[D,T_0] = iT_0\,\quad [D,T_1] = iT_1\label{ND}\,.
\end{align}
Rewriting such algebra in terms of light cone generators $P_{\pm} \equiv T_0\pm T_1$ and $R_{\pm} \equiv-1/2( N\pm D)$ it os easy to see that we can make the following identification with the generators of the algebra \eqref{axbalg}: $P_+=P$ and $R_+=R$. Thus representations of the $ax+b$ group are directly related to the one particle states of a massless field on Minkowski and Rindler spaces. As it is well known, the operator $aN$, where $a$ is the Rindler observer acceleration scale, has the physical meaning of the Rindler space Hamiltonian, while the operator $aD$ generates Rindler spatial translations. It follows that to describe physics as seen by the accelerated observer moving with acceleration $a$, we have to replace the operator $R$ with $aR$, the latter having the canonical dimension of inverse length. Tracing the resulting changes down to the equation \eqref{thermak}, we discover that the physical temperature of thermal distribution becomes $T=a/2\pi$, which is exactly Unruh temperature.

The derivation we presented unveils a fundamental group theoretic structure underlying the thermodynamic properties of quantum systems in the presence of partitions induced by different choices of observables. This evidences the universal character of such effects and suggests that they should be a common feature of {\it any} quantum system whose fundamental symmetries are described by the affine group. We hope that such new look on the Unruh effect based on simple elements of representation theory can provide further valuable insight on the puzzling aspects of quantum effects near the horizon and the fate of unitarity in the evolution of a black hole.

\section*{Acknowledgment}
For JKG this work is supported by the National Science Center, project number
2017/27/B/ST2/01902.

\end{document}